%Paper: 9204012
%From: sethna@nordita.dk (James P. Sethna)
%Date: Thu, 23 Apr 92 10:12:10 +0200

%%%%%%%%%%%%%%%%%%%%%%%%%%%%%%%%%%%%%%%%%%%%%%%%
%                                              %
% The tex file with text and figure captions   %
% begins here.  Four postscript files, for     %
% figures 1 -- 4, follow; they are             %
% separated by headings like this one.         %
% These should be separated into their own     %
% files and then printed separately.           %
%                                              %
% If you need the macro packages jnl,          %
% reforder, and eqnorder, send e-mail to       %
% sethna@helios.tn.cornell.edu                 %
%					       %
%%%%%%%%%%%%%%%%%%%%%%%%%%%%%%%%%%%%%%%%%%%%%%%%

\input jnl
\input reforder
\input eqnorder

\title Tweed in Martensites: A Potential New Spin Glass
\author James P. Sethna \footnote{$^\ddagger$}{Permanent Address: Laboratory %
of Atomic and Solid State Physics, Cornell University, Ithaca, New York %
14853-2501}
\affil Laboratory of Applied Physics, Technical University of Denmark, %
DK-2800 Lyngby, DENMARK, and NORDITA, DK-2100 Copenhagen \O, DENMARK
\author Sivan~Kartha, Teresa~Cast\'an,\footnote{$^*$}{Permanent address: %
Departament d'Estructura i Constituents de la Mat\`eria, Facultat de %
F\'isica, Universitat de Barcelona, Diagonal 647, 08028 Barcelona.}
and James~A. Krumhansl,\footnote{$^\dagger$}{Present Address: 515 Station Rd.,%
 Amherst, Massachusetts 01002}
\affil Laboratory of Atomic and Solid State Physics, Cornell University, %
Ithaca, New York 14853-2501

\abstract

We've been studying the ``tweed'' precursors above the martensitic transition
in shape--memory alloys.  These characteristic cross--hatched modulations
occur for hundreds of degrees above the first--order shape--changing
transition.  Our two--dimensional model for this transition, in the
limit of infinite elastic anisotropy, can be mapped onto a spin--glass
Hamiltonian in a random field.  We suggest that the tweed precursors
are a direct analogy of the spin--glass phase.  The tweed is intermediate
between the high--temperature cubic phase and the low--temperature martensitic
phase in the same way as the spin--glass phase can be intermediate between
ferromagnet and antiferromagnet.


\noindent Submitted for publication in Physica Scripta.

\endtopmatter

\noindent
\nobreak

The field of disordered systems has been a rich and fascinating branch
of condensed matter physics in the last few decades.  New ideas,
language, and techniques have been developed in order to study problems
with intrinsic randomness in their Hamiltonians.  Much of the attention
has been on relatively new and exotic systems: spin glasses in dilute magnetic
systems, charge density waves in complex one--dimensional materials,
localization in doped semiconductors.  One of the promises of the field
is to treat the real, dirty world of materials.  So long as our industrial
colleagues refrain from making their goods with perfect crystals, the
study of disordered systems ought to have ramifications in old--fashioned,
practical systems.

One of the old problems in materials physics is that of the ``central peak''.
First--order transitions traditionally have no precursors: water looks
like water until 0$^\circ$C, at which point it suddenly becomes plain ice.
Transitions between different crystalline phases, in contrast, often
have rather large precursor effects.  In neutron scattering, there
is often a large, quasielastic central peak, representing either static
or rather long--lived fluctuations in positions.  In the martensitic
phase transitions (described more fully below) the central peak
may be understood to be associated with a tweed morphology in TEM
micrographs.  The tweed,\refto{Tanner} so called because of its characteristic
irregular fabric--like cross--hatched pattern, has stripes aligned along
$\langle 110 \rangle$ directions of widths a few atomic spacings.
Tweed occurs for tens to hundreds of degrees above the transition temperature.

First, we'll review what a martensitic transformation is.  Second, since
martensitic materials have high elastic anisotropies,
we'll take a limit of infinite elastic anisotropy to explain the
tweed morphology.  Third, we'll introduce disorder, and show that
the tweed regime can be understood (in a two--dimensional model) as
a spin--glass phase intermediate between the high and low--temperature
phases.
Thus, the community's large investment of work on spin glasses is seen
to have direct application to a substantial problem in traditional
metallurgy.  This is a simpler presentation of recently published
work.\refto{Sivan}

Martensitic phase transformations were first studied in carbon steels.
Typically, they involve a cubic high--temperature phase
which abruptly changes shape, elongating in one direction
and contracting in the other two, with possible intracell,
atomic shufflings.  The resulting low-temperature phase is
usually left with a complicated morphology of micron--sized plates,
which in turn can be composed of many parallel twinned regions.
The simplest examples of martensitic transitions
are the bcc $\to$ fcc transition or bcc $\to$ hcp transitions seen in
many metals and alloys on cooling.  The martensitic transformations of interest
to us have (1)~little volume change, and (2)~no long--range diffusion
during the transition.  Large
changes in volume prevent the nice elastic accommodation that the
plates represent: crystals undergoing such transitions may in fact
shatter from the resulting strains.  The martensitic transformations
in carbon steels have too much volume change to be described by our theories.
If long--range diffusion is
needed in a transition, it usually proceeds rather slowly: in contrast, the
growth of a martensitic plate can occur at the speed of sound,
and emits a cracking noise.

We model these materials with a two--dimensional simulation.
We represent each atomic cell by a quadrilateral: the two types
of atoms in the simulation are represented by quadrilaterals
with two different energies as a function of shape.  The high
temperature phase in our simulation is square (grey in figures 1 and~2),
while the low temperature phase has two variants: tall-and-thin (black), and
short-and-fat (white).  The martensites we're interested in don't change
shape drastically: a couple of percent stretch along one axis is typical.
(The shape change in the figures is exaggerated by a factor of 15.)

Because of this, these materials have transitions which are largely reversible.
In figure~1, we see a typical martensitic region.  One sees domains
of both variants, separated by twin boundaries along (in this case) the
diagonal stretching from lower left to upper right.  Notice that the
region as a whole has preserved it's square shape: the twin boundaries
arrange themselves so as to prevent long--range strain
from building up.  This is for precisely the same reasons that magnetic
domains form in ferromagnets: the long range elastic (magnetic) fields
introduced from one growing variant lower the energy of formation of the
other variant in neighboring regions.

There are three types of materials in which tweed has been seen.  First (and
most recently), it has been seen in doped
YBaCuO\refto{schmahl, Moss, Semenovskaya}, where
our two--dimensional model is particularly appropriate.  YBCO goes through
a tetragonal $\to$ orthorhombic phase transition somewhat above its
superconducting transition.  If one views it in a funny basis, this
transition is a square $\to$ rectangular transition in the copper--oxygen
planes.  When cobalt or aluminum is doped for copper, it apparently
substitutes onto the chains, producing a substantial drop in the
superconducting transition temperature and a tweed--like modulation.
Second, it has been seen in the A15's\refto{onozuka}, which also
go through a martensitic transition near the superconducting transition.
Third (and the way we got interested), it has been seen in the shape--memory
al\-loys.\refto{oshima, robway, Shapiro, shapiroyang, seto, schryverstan}

In passing, we can use figure~1 to provide a simple illustration of
the shape--memory effect.  Imagine the large square sample as some object (say
a teapot) formed out of the high--temperature austenite phase.  We've
passed through the phase transition, yet the region is roughly
square: twin boundaries have formed to keep the material
macroscopically the same shape (square, or teapot).  In the martensitic
phase, though, the material is much softer, even though the single--crystal
elastic constants are stiffer.  To plastically deform the
teapot, one needn't break any interatomic bonds.  One need only push
the twin boundaries around!  If one takes the sample in figure~1 and
stretches it vertically, the twin boundaries will shift to shrink the
white regions.  One can crumple the teapot without undue force, introducing
no changes in the underlying cubic lattice.  Upon reheating,
the material returns to the
cubic phase.  The twin boundaries, wherever they have moved, will
disappear, and the material will reform into its original configuration.
The teapot miraculously uncrumples!  Shape--memory alloys are used
industrially for thermal switches (often in automobiles), and for
pipe fittings.\refto{book}  They've been used also in more exotic
applications, such as satellite antennae which automatically unpack
themselves, robotic muscles which flex upon resistive heating, and
small bars which bust up rocks\refto{rocks} in golf courses.

Figure~2 shows another local low--energy state of our model, here in the
tweed temperature range.  Notice the patchy cross--hatched diagonal
checkerboard pattern.  There are long--ranged correlations in
the two diagonal directions, but only short--ranged correlations
horizontally and vertically.  Indeed, the dark and light stripes
extend in many cases through our whole sample.

Let's define the order parameter for our martensitic transition
to be the net stretch in the horizontal direction:
$$\phi = \epsilon_{xx} - \epsilon_{yy}, \eqno(phi)$$
where the strain field $\epsilon_{ij} = (\partial_i u_j + \partial_j u_i)/2$
and $u(x,y)$ is the displacement field (final position minus initial
position, labeled by the initial position $x,y$).
(We have in the analysis ignored the ``geometric nonlinearity''
$\partial_i u_\ell \partial_j u_\ell$, which can be important if there
are large rotations in the problem.  Our unconstrained simulations do implement
complete rotation invariance.) Thus, $\phi > 0$ in short-and-fat
regions, $\phi = 0$ in the square phase, and $\phi < 0$ in the tall-and-thin
regions.

Now, writing a free energy in terms of strain fields
is tricky.  Indeed, most configurations of $\epsilon(x,y)$
are illegal!  Unless
$${\bf Inc}(\epsilon) = \nabla \times (\nabla \epsilon)^T \eqno(Inc) $$
is zero, the strain field describes a material with embedded line
dislocations.\refto{Baus}  Doing any calculations in terms of strain
fields must either add a term for the energy of dislocations, or must
put in some kind of Lagrange multiplier to forbid $\epsilon$ from developing
an {\bf Inc}.
Working in terms of $\phi(x,y)$ would seem even worse.  Since the components
of $\epsilon$ are not independent \(Inc), most configurations of
$\phi=\epsilon_{xx}-\epsilon_{yy}$ will necessarily involve other
kinds of strain.  If we allow $\phi$ to vary freely, we must ``integrate
out'' the other elastic degrees of freedom, leaving us with a free
energy with long--range forces.

There is another option, which is suggested by the large elastic anisotropy
characteristic of martensites.  Because the high temperature phase is cubic,
it has three independent elastic constants (rather than two for an
isotropic material).  The elastic constant C$'$ which resists deformations into
rectangular shapes is much smaller\footnote{$^\dagger$}{Actually,
the measured elastic constants in the tweed and martensitic regime
will have contributions from rearrangements of the domains, as in the
A15's\refto{Testardi}.  How much the tweed is due to the lowering of
C$'$ and how much the lowering of C$'$ might be due to the response
of the tweed morphology is a subject we plan to explore.}
than the elastic constant which resists bulk compression
($\epsilon_{xx} + \epsilon_{yy} \ne 0$) and the one which resists diagonal
strain $(\epsilon_{xy} \ne 0$);  we let these two elastic constants
go to infinity, thus constraining the system to forbid any deformation except
rotations, translations, and rectangular stretches (measured by
$\phi$).

These constraints profoundly restrict the configurations.  A straightforward
com\-pu\-ta\-tion\refto{Sivan} shows that the two dimensional configurations
that are allowed are precisely those which can be written as a sum of
two one--dimensional strain modulations\refto{Ericksen}
$$\phi(x,y) = \phi^+(x+y) + \phi^-(x-y), \eqno(Constrained)$$
$\phi^+$ describing a modulation along the $(x,y)$ direction (lower left to
upper right), and $\phi^-$ describing a modulation along the $(x,-y)$
direction.  (Equation \(Constrained) is derived by first showing that the
displacement ${\bf u(x,y)}$ can be written as a sum of two one--dimensional
displacement fields, and then finding the strains.)

Now, it's been known for a long time that large elastic anisotropies
are associated with strain fields which extend long distances along the
diagonals.  We are encouraged for three reasons to think that the
infinite anisotropy limit is a natural starting point for studying tweed.
First, the form \(Constrained) describes the behavior of our simulation well.
The displacement variation in figure~1, for example,
is nearly purely of the $\phi^-$
form.  The variation of figure~2 is well described as a superposition
\(Constrained).  Second, it plausibly describes the experimental
tweed morphology.  The experimentalists agree with us that tweed is composed
of $\{110\}\langle 1{-1}0 \rangle$ shears, and that it looks
cross--hatched.\footnote{$^*$}{Actually, the experimentalists haven't
for sure decided that the tweed really is a superposition.
Since TEM measures through a rather thick slab, there is
no direct evidence that the observed cross-hatched patterns aren't
plain old twins pointing in different directions at different depths.
In two dimensional systems, at least, we unambiguously predict
modulations in both $\langle 11 \rangle$ directions at once.}
Third, it describes the modulation patterns seen in earlier and
later simulations, with different driving mechanisms for the
modulations.\refto{Khachaturyan, Semenovskaya} Thus, in
the limit of infinite elastic anisotropy, cross--hatched patterns
become the only allowed patterns.

So far, we have explained how the tweed morphology is a natural modulation
for materials with large elastic anisotropy.  We now must address why
these materials choose to modulate at all.  Experimentally, in
closely related systems, neutron scattering experiments in nearly
pure materials have illuminated the important role of disorder.\refto{Heiming}
Moreover recent simulations have found tweed in a disordered alloy but not
in an ordered one.\refto{Clapp}  We consider here the disorder
introduced by concentration variations.

Because the shape memory
alloys are alloys, static statistical concentration variations provide
an intrinsic source of randomness.  First, the martensitic
transition temperature varies drastically with concentration
$\eta$ in most of these systems: in
Fe$_{1-\eta_0}$Pd$_{\eta_0}$, the fcc$\rightarrow$fct transition is
at room temperature for $\eta_0 = 0.29$, and drops to 0K for $\eta_0 = 0.32$.
This will produce a free energy difference
$F_A(\eta(x,y)) - F_M(\eta(x,y))$ (or more concisely, $(F_A - F_M)(\eta)$)
which varies with position.  At high temperatures, this will always
be negative, at low temperatures always positive, but near the transition
temperature at the average concentration, this difference will vary in
sign from place to place.  Second, the local configurations of Fe and Pd atoms
can introduce local distortions, favoring one or another of the martensitic
variants.\refto{robway, Moss, Clapp}  This will produce a free energy
difference
$(F_--F_+)(x,y)$ which will depend on gradients of $\eta$.\refto{Sivan}

The local elastic free energy can in general have three minima: two at
$\pm \phi_0$ for the two martensitic variants, and one at $\phi = 0$
for the austenite.\footnote{$^\ddagger$}{$\phi_0$ will generally depend
upon temperature, and at high enough temperatures there will be no martensitic
minima.  This just contributes to the effective temperature dependence of the
bonds, described below.}
If we define $i=x+y$ and $j=x-y$,
then each $\phi^+_i$ and each $\phi^-_j$ will represent a contribution to
the local order parameter for all the sites along some diagonal.  The order
parameter at a certain site $(x,y)$ will be given by the sum of $\phi^+_i$
and $\phi^-_j$ for appropriate $i$ and $j$.  In our constrained system
\(Constrained), it's possible to arrange to have the local order parameter
modulate solely between the three states by letting $\phi^+$ and $\phi^-$
each take on two possible values, $-\phi_0/2$ or $+\phi_0/2$.
If we then think of the $\phi^+_{i}$ and
$\phi^-_{j}$ as spins that either point down or up, then we have a
convenient spin representation for conceptualizing the tweed problem.
Either of the two antiparallel configurations produces an undeformed,
square, austenite region ($\uparrow\downarrow$ or $\downarrow\uparrow$ give
$\phi=0$), and the parallel configurations produce
the two rectangular martensitic variants ($\uparrow\uparrow$ gives $+\phi_0$,
$\downarrow\downarrow$ gives $-\phi_0$).

The tweed model, in spin language, is precisely equivalent to an infinite
range spin glass.  The spins in the spin glass are analogous to the diagonals
which connect the sites in the martensite.  In the spin glass, each spin is
coupled to every other spin by a particular bond with a random sign
and strength.  In the tweed, each diagonal is ``coupled'' to every other
diagonal by a particular site, which prefers austenite ($\phi=0$,
$\uparrow\downarrow$ or $\downarrow\uparrow$) or martensite ($\phi=\pm \phi_0$,
$\uparrow\uparrow$ or $\downarrow\downarrow$) at random.
In each case, the couplings represent the frozen--in disorder.

Figure~3 illustrates the analogy between the two systems.
Each letter ({\bf A,F}; {\bf A,M}) represent a
terms in the energy: the spins, squares and rectangles show
one of the possible metastable configurations of each of the systems.
Both the spin glass and the tweed
are frustrated: not all the connections can be made happy at the same time.
In both cases, any closed loop with an odd number of {\bf A}'s is frustrated:
one of the bonds in the spin glass must be broken, and one of the
sites in the martensite must locally be in the wrong phase.

To complete the analogy, one must notice three more correspondences.
First, just as the effective bond strength $J_{ij}$ between two
diagonals is given by $(F_A-F_M)(\eta)$ at the corresponding $(x,y)$,
there is a contribution to the effective ``magnetic field'' $H_{ij}$
on the two diagonals given by $(F_--F_+)$.  Second, the free
energy difference $F_A-F_M$ is strongly temperature dependent.
In the spin glass, the Hamiltonian really represents all the degrees of
freedom:
the parameters are the bare ones, and do not depend on temperature
(although they might depend, for example, on pressure).  In our
problem, the martensitic transition is (probably) driven by vibrational
entropy: the phase space of small oscillations is not represented
by the degrees of freedom $\phi(x,y)$ we're including in our free energy.
Tracing over these small vibrations makes the bonds $J_{ij}$ in our
free energy temperature dependent: roughly speaking,
$-J_{ij}(T) = -J_{ij}^0 + L T$, where $L$ is the vibrational latent
heat per site.  Third,
one must notice that every lower-left--upper-right diagonal crosses
every upper-left--lower-right diagonal.  The interactions are of
infinite range!\footnote{$^\dagger$}{This happens only for infinite
elastic anisotropy, of course.}  Table~I shows the correspondences,
up to things like factors of 2 and $\phi_0^2$.

\bigskip
\centerline{\bf Table I: Correspondence between Spin Glass and Tweed
Parameters}
\halign to \hsize{\hskip1truecm#\hfill&\hfill#\hfill&\hfill#\hfill&\hfill#%
\hfill&\hfill#\hfill&\hfill#\hfill&\hfill#\hfill&\hfill#\hfill\cr
Spin Glass~~~&$i$&$j$&$s_i$&$s_j$&$J_{ij}$&$H_{ij}$&$T$\cr
Tweed&~~~$x+y$~~~&~~~$x-y$~~~&~~~$\phi^+$~~~&~~~$\phi^-$~~~&~~~$F_A-F_M$~~~&%
{}~~~$F_- - F_+$~~~&~~~$\langle -J \rangle_{ij}/L$~~~\cr}
\bigskip

We can now write a spin--glass like Hamiltonian for the martensitic
tweed system:
$${\cal F} = \sum_{i,j} -J_{ij} \phi^+_i \phi^-_j - H_{ij}(\phi^+_i +
\phi^-_j).
								\eqno(H)$$
Every site $i$ interacts with every site $j$: it corresponds to an
infinite range spin glass\refto{sk}, on which we expect replica
theory to give an exact solution.\refto{replica}  Unlike the SK
model, there are two classes of spins, each of which interacts only
with the other class: it is a {\it bipartite infinite--range spin
glass}.\refto{Korenblit}

Figure~4 shows the phase diagram we expect for the free energy equation \(H),
setting the random field term to zero.  What do I mean by ``expect''?  While
this particular model has never been completely solved, we can guess
the answer from the partial solutions and from the behavior of the
SK model.  Korenblit and Shender\refto{Korenblit} solved this model in the
replica symmetric approximation.  This is known to give the correct state
in the paramagnetic, ferromagnetic, and antiferromagnetic phases,
and the right phase boundaries for these phases.  The phase diagram
they found was exactly that of the SK model with varying fractions
of ferromagnetic bonds, except that the bipartite model phase diagram
was doubled by reflection through $\langle J \rangle = 0$.  The
replica--symmetry breaking solution to the SK model\refto{replica}
has a vertical phase boundary between the spin--glass phase and
a glassy phase with long--range order (called the magnetized spin--glass,
or MSG phase).  We expect, thus, to find vertical phase boundaries
in the bipartite model too: hence figure~4.  The problem with
the random field has been studied too.\refto{Pirc}

What can we see from this phase diagram?  Tweed is a spin--glass
phase intermediate between the two ordered phases.  The
antiferromagnetic phase on the right has the two sublattices pointing in
opposite directions: apart from thermal fluctuations, the entire region has
$\phi = \phi^+ + \phi_- = 0$.  This is the high--temperature square
austenite phase for our model.  The ferromagnetic phase on the left
has the two sublattices pointing in the same direction: the region
has transformed entirely into the low--temperature martensitic phase.
The paramagnetic phase on the top also has square symmetry on average, but is
a thermal mixture of square regions together with both rectangular variants.
Conceivably, this could represent the austenite phase too.  (Could
bcc not be the familiar lattice plus small oscillations, but really sometimes
be an equal mixture of various other variants?  Recent experiments
on zirconium suggest that this actually might be the case!\refto{Heiming})
Finally, the spin--glass phase corresponds to tweed.  It is a frustrated
attempt to accommodate the concentration fluctuations: a patchy, glassy
mixture of square and rectangular regions.  Because our effective
bonds are temperature dependent, heating the physical martensite
moves us along the dotted line in the figure: both the average concentration
of antiferromagnetic bonds and the temperature of the ``spin'' system
increases.

Young people giving their first talks often fear that someone will ask
the key questions which they know will expose the gaping holes in their
theories.  I would like to conclude with my attempt to answer
three such questions.

Q. {\it I thought the martensitic transition was first order.  Also, does
your theory work in three dimensions?}

A.  Indeed, the Ising spin--glass transition is second order.  I guess
in two dimensions ({\it i.e.}, extremely thin high--T$_c$ samples) we
do predict that the first--order martensitic transition will be replaced
by two second--order transitions in the presence of sufficient disorder.
(Imry and Wortis\refto{Imry} have discussed the effects of
random--field disorder, which can destroy the transition entirely.)
In three dimensions, the system is clearly not an Ising spin glass.
First, there are at least three martensitic variants: the order
parameter $\phi$ becomes a two component object with three local
minima stretching in the three orthogonal directions.  Second,
in the limit of infinite elastic anisotropy, there are six scalar
one--dimensional functions corresponding to $\phi^\pm$.  Even changing
from an Ising to a Potts glass\refto{Potts} makes for a first--order
spin--glass transition (albeit with no latent heat).  What kind of
transitions we will find once we identify the right spin model for
3D martensites is completely open at this point.

Before mindlessly
generalizing our work from two to three dimensions, we want to
get a better physical picture of what is important.  We're
exploring effective--medium simulations with Karsten Jacobsen
and TEM experiments with John Silcox in order to
get a handle on the impurity couplings, and on the important long--range
correlations.  If I have to guess, I'd say that there is a better
chance of a real spin--glass phase in 3D.  In 2D, tweed is a true
phase only in mean--field (infinite elastic anisotropy): in 3D, a
sufficiently frustrated system ought to have a true glassy phase.

Q. {\it Why haven't you taken parameters from the 2D high--T$_c$ materials,
instead of the uncontrolled extrapolation from FePd?  Also,
do you have any predictions for strontium--doped LaCuO?}  (Mac Beasley
actually asked the second question, and pointed out some experimental
data that we should pursue.)

A.  We were, oddly enough, more interested in the shape--memory alloys,
and it took us a while to realize how hard 3D was going to be.  High T$_c$ is
next on the agenda.

Q. {\it Do you have any experimental predictions?}

A. We don't have definite predictions for the statics in 3D.  We'd
really like to predict a diverging nonlinear elastic susceptibility.
As one crosses the phase diagram in figure~4 from the paramagnetic
phase to the tweed spin--glass phase, the nonlinear susceptibility
diverges.  That is, if we put an external force $F$ which stretches
the sample coupling to $\phi$, and measure the response
$\phi = \chi_1 F + \chi_3 F^3$, then $\chi_3$ will diverge at the
phase boundary.  There are three problems, though.  First, the
susceptibilities which diverge along the side phase boundaries
of the tweed
apparently aren't elastic constants.\refto{replica}  Unless
the high--temperature phase is a melt of the various low--temperature
variants (as discussed above), so the austenite corresponds to a paramagnet,
we don't get an effect.  Second,
in two dimensions the mean--field transition will be rounded whenever
the elastic anisotropy isn't infinite.  Third, the predictions in
three dimensions are likely to be quite different.  If I were an
experimentalist, I'd certainly try looking for phase boundaries, using
elastic probes.  We can't say yet, though, exactly what to look for.

The dynamical predictions are much more straightforward.  All of
the glassy systems have interesting, slow dynamics.  Even if our
spin--glass identification is wrong, and random--field effects
dominate the problem, there should be memory, hysteresis, and
logarithmic decays in response functions.  While this may seem
old hat, in metallurgy this ought to be more exciting.  I'm not
saying, of course, that slow elastic relaxation in dirty metalurgical
systems is news: what ought to be exciting is that there may be
interesting, quantitative science to be done with the dirt.

\vfill\break

\centerline{\bf Figure Captions}

\bigskip
\noindent {\bf Figure~1: Martensitic Twins}.
A typical metastable state of our two--dimensional
model, with parameters set to be in the low--temperature
rectangular martensitic phase.  The free energy is of the same form
as given in our earlier publication\refto{Sivan}, except that we've
added energy terms for bulk compression and diagonal strain ({\it i.e.}, this
is an unconstrained simulation).  Parameters are taken from the
Fe$_{0.7}$Pd$_{0.3}$ shape memory alloy, with each quadrilateral representing
one atom.  Naturally, significant simplification was made in going from
three dimensions to two.  In addition, we don't know yet the coupling to
impurities.  Details will be published in a longer form.\refto{Future}

Notice the twin boundaries between the two variants: they lie along
the $\langle 11 \rangle$ diagonals.  This nearly one--dimensional
variation is of the
form $\phi^-(x-y)$ as described in the text, equation \(Constrained).
Notice that twin boundaries have arranged to keep
the overall shape of the domain the same as that of the original sample:
a large chunk of one variant would have produced intolerable strain
in the surrounding matrix.

\bigskip
\noindent {\bf Figure~2: Tweed}.
Another metastable state, at parameters
set in the tweed regime.  Notice the patchy tweed pattern, looking
like a superposition of stripes along the two diagonals.  Grey is
the high--temperature square phase, black is tall-and-thin, white
is short-and-fat.  Different quenches typically
land in different metastable states: the system is glassy.

\bigskip
\noindent {\bf Figure~3: Frustration}.  Our model of martensites is
frustrated in precisely the same way as are spin glasses.

{\bf Spin Glass}.  On the right, we have three spins which can either
point up or down.  They are connected by bonds which prefer parallel
({\bf F}erromagnetic) or antiparallel ({\bf A}nti\-fer\-ro\-mag\-net\-ic)
alignment.  In a spin glass, there are many such spins, and
the bonds are chosen {\bf F} or {\bf A} at random.  If we start
at the lower left corner of the triangle, starting (arbitrarily) down,
and move around clockwise, we can satisfy each bond in turn.  However,
for any loop with an odd number of antiferromagnetic bonds, one bond
must be broken (in its high-energy state): here the ferromagnetic bond
connecting the bottom two spins is not satisfied.

{\bf Tweed.}  On the left, we have four regions in the material, forming
a diamond.  Starting at the bottom, we have a region where the local
alloy concentration makes the square, {\bf A}ustenite phase lower in
energy.  In the limit of infinite elastic anisotropy, the order
parameter at the bottom site is the sum of two contributions, one
for each diagonal.  By choosing the contribution for the lower left
diagonal $+\phi_0/2$ (represented by a plus spin $\uparrow$), and
the contribution for the lower right diagonal $-\phi_0/2$ ($\downarrow$),
we satisfy the local free energy in the bottom site: $\phi = 0$,
represented by the square box surrounding the {\bf A}.  Again, moving
clockwise, we can satisfy the free energy in the left--hand site
(which prefers {\bf M}artensite) by choosing $+\phi_0/2$
($\uparrow$) for the upper left diagonal, leading to a net order
parameter $\phi_0$ and a short--fat rectangular deformation.
We can continue this process through the
top site, but when we reach the final, right--hand site, the order
parameter is already determined to be $0 = + \phi_0/2 - \phi_0/2$
($\uparrow\downarrow)$: the local free energy, which prefers {\bf M},
is not satisfied.

\bigskip
\noindent {\bf Figure~4: Phase Diagram}.  The phase diagram
expected\refto{Korenblit}
for the infinite--range bipartite spin-glass model, equation \(H).
A given sample will traverse a sloped path in this phase diagram
as temperature is increased, because of the temperature dependence
in the effective coupling.  It will pass from the martensitic phase
through the tweed regime into the austenite phase, as shown by the
arrow.  The two
magnetized--spin--glass phases (FMSG and AMSG) are supposed to have
glassy dynamics and metastability, even though they have long--range order.
We ignore here the random field term $H_{ij}$.

\vfill\break
\centerline{\bf Acknowledgments}

We acknowledge the support of DOE Grant \#DE-FG02-88-ER45364.  T.~C.
wishes to thank the Laboratory of Atomic and Solid State Physics for
their hospitality, and the D.~G.~I.~C.~Y.~T. (Ministry of Education,
Spain) for financial support. J.~P.~S. would like to thank the
Technical University of Denmark and NORDITA for support and hospitality.

\references

\refis{Tanner} L.~E. Tanner, \journal{Phil. Mag.}, 14, 111, 1966.

\refis{Sivan} Sivan~Kartha, Teresa~Cast\'an, James~A. Krumhansl, and
James~P.~Sethna, \prl 67, 3630, 1991.

\refis{Future} Sivan~Kartha, Teresa~Cast\'an, James~A. Krumhansl, and
James~P.~Sethna, to be published.

\refis{rocks} Made by Tokin Corporation in Sendai, Japan, see
{\sl Science News}, p. 339, December 14, 1991.

\refis{book} For a good introduction to shape--memory alloys and their
applications, see {\it Engineering Aspects of Shape Memory Alloys},
edited by T.~W. Duerig, K.~N. Melton, D.~St\"ockel, and C.~M. Wayman,
Butterworth--Heinemann, London, 1990.

\refis{Baus} M.~Baus and R.~Lovett, \prl 65, 1781, 1990.

\refis{Moss} X.~Jiang, P.~Wochner, S.~C. Moss, and P. Zschack,
\prl 67, 2167, 1991.

\refis{onozuka} T.~Onozuko, N.~Ohnishi and M.~Hirabayashi, {\sl Met. Trans.
A},{\bf 19A}, 797 (1988).

\refis{schmahl} W.~W. Schmahl, A.~Putnis, E.~Salje, P.~Freeman,
A.~Graeme-Barber, R.~Jones, K.~K. Singh, J.~Blunt, P.~P. Edwards, J.~Loram,
and K.~Mirza, {Phil. Mag. Let},{\bf 60}, 241, (1989).

\refis{robway} I.~M. Robertson and C.~M. Wayman, {\sl Phil. Mag. A}, {\bf 48},
421, 443, and 629, (1983).

\refis{oshima} R.~Oshima, M.~Sugiyama and F.~E. Fujita, {\sl Met. Trans}, {\bf
19A}, 803, (1988).  S.~Muto, R.~Oshima, F.~E. Fujita, {\sl Acta Metall.
Mater.}, {\bf 38}, 684, 1990.  S.~Muto, S.~Takeda, R.~Oshima, F.~E. Fujita,
{\sl J. Phys: Condens. Matter}, {\bf 1}, 9971, 1989.

\refis{schryverstan} D.~Schryvers and L.~E. Tanner, {\sl Mat. Sci. Eng.}
{\bf A127}, (1990).  D.~Schryvers, L.~E. Tanner and S.~Shapiro, {\sl
Ultramicroscopy}, {\bf 32}, 241, (1990).

\refis{Shapiro} S.~M. Shapiro, Y.~Noda, Y.~Fujii, and Y.~Yamada,
\prb 30, 4314, 1984.

\refis{shapiroyang} S.~M.~Shapiro, B.~X.~Yang, G.~Shirane, Y.~Noda and
L.~E. Tanner, {\prl 62, 1298, 1989}, S.~M. Shapiro, J.~Z. Larese, Y.~Noda,
S.~C. Moss and L.~E. Tanner, {\prl 57, 3199, 1989}

\refis{seto} H.~Seto, Y.~Noda, and Y.~Yamada, {\sl J. Phys. Soc. Japan},
{\bf 59}, 965, (1990).

\refis{Testardi}  L.~F. Testardi and T.~B. Bateman, {\sl Phys. Rev.} {\bf 154},
402 (1967), cf. also L.~F. Testardi, \rmp 47, 637, 1975.

\refis{Heiming} A.~Heiming, W.~Petry, G.~Vogl, J.~Trampenau, H.~R.
Schober, J.~Chevrier, and O.~Sch\"arp, {\sl Z. Phys B} {\bf 85}, 239 (1991),
and \prb 43, {10933, 10948, and 10963}, 1991.  We thank P-A.~Lindg\aa rd
for explaining this to us.

\refis{Ericksen} These solutions have been noted for two dimensions by J.~L.
Ericksen ({\sl Int. J.  Sol.  Struct.}, {\bf 22}, 951, (1986)) and A.~E.
Jacobs (\prb 31, 5984, 1985.)  Our result also applies, {\it mutadis
mutandis}, to three dimensions.

\refis{Semenovskaya} S.~Semenovskaya and A.~G. Khachaturyan,
\prl 67, 2223, 1991.

\refis{Khachaturyan} A.~Khachaturyan, {\sl Theory of Structural
Transformations in Solids},
(Wiley-Inter\-sci\-ence, 1983).

\refis{Clapp} C.~Bequart, P.~C. Clapp, and J.G. Rifkin, in
{\it Kinetics of Phase Transitions}, edited by M.~E. Thompson
{\it et. al}, MRS Symposia Proceedings No.~25 (Materials Research
Society, Pittsburgh, 1990).

\refis{sk} D.~Sherrington and S.~Kirkpatrick, \prl 35, 1792, 1975.

\refis{replica}
G.~Parisi, \journal{Physics Lett. A}, 73, 203, 1979, \prl 43, 1754,
1979, {\sl J. Phys. A} {\bf 13}, L115, 1101, 1807 (1980),
M.~Mezard, G.~Parisi, and M.~A. Virasoro, {\it Spin Glass
Theory and Beyond,} World Scientific, Singapore, 1987, and
K.~Binder and A.~P. Young  \rmp 58, 801, 1986.

\refis{Korenblit} I.~Ya Korenblit, E.~F. Shender, {\sl Zh. Eksp. Teor. Fiz.},
{\bf 89}, 1785, (1985).  H.~Takayama, {\sl Prog. Theor. Phys.}, {\bf 80},
827, (1988).

\refis{Pirc} R.~Pirc, B.~Tabic and R.~Blinc,{\sl Phys. Rev. B}, {\bf 36},
8607, (1987).

\refis{Imry} Y.~Imry and M.~Wortis, \prb 19, 3580, 1979.

\refis{Potts} D.~J. Gross, I.~Kanter, and H.~Sompolinsky, \prl 55, 304, 1985,
D.~Thirumalai and T.~R. Kirkpatrick, \prb 38, 4881, 1988 and references
therein.

\endreferences

\enddocument
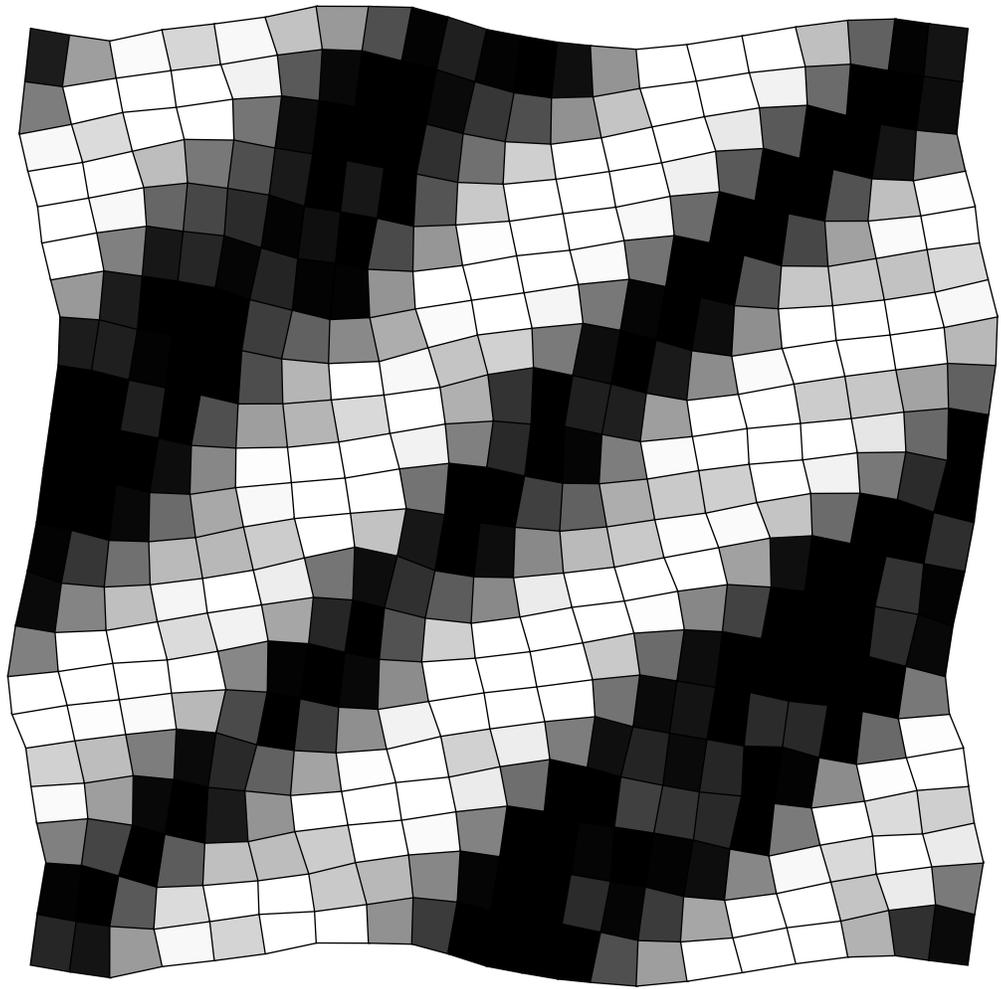
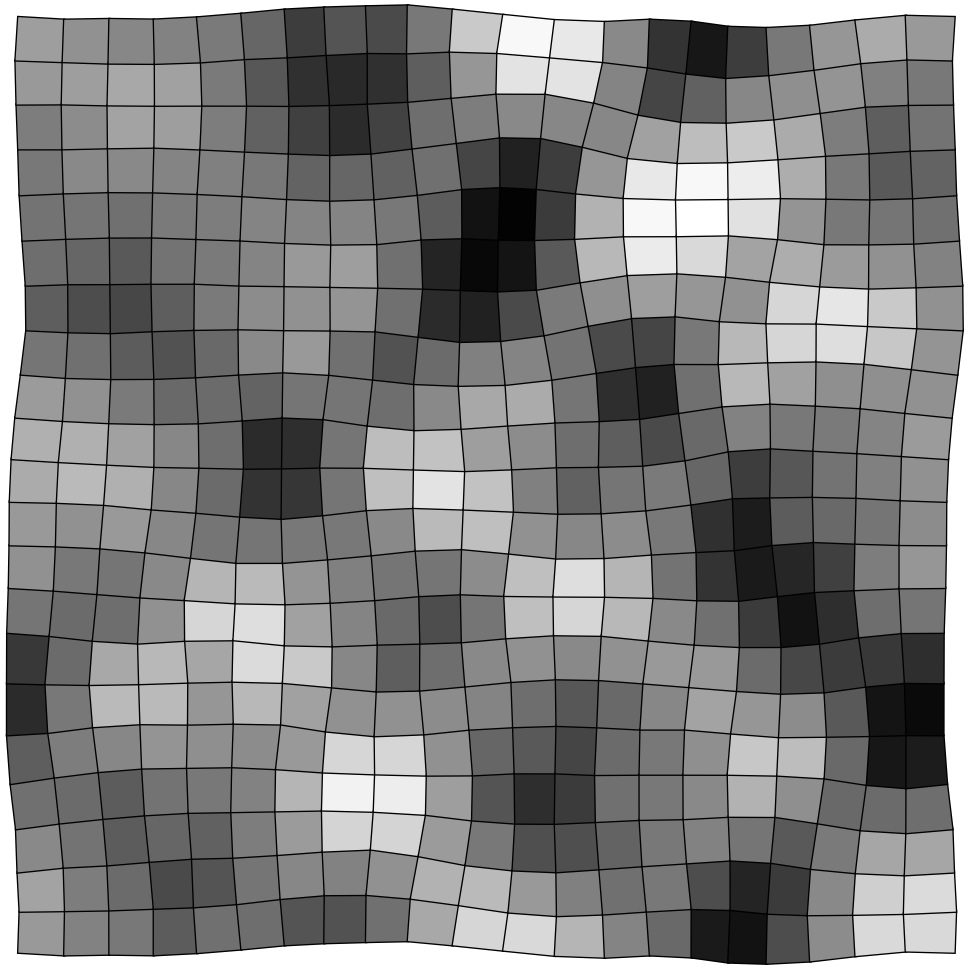
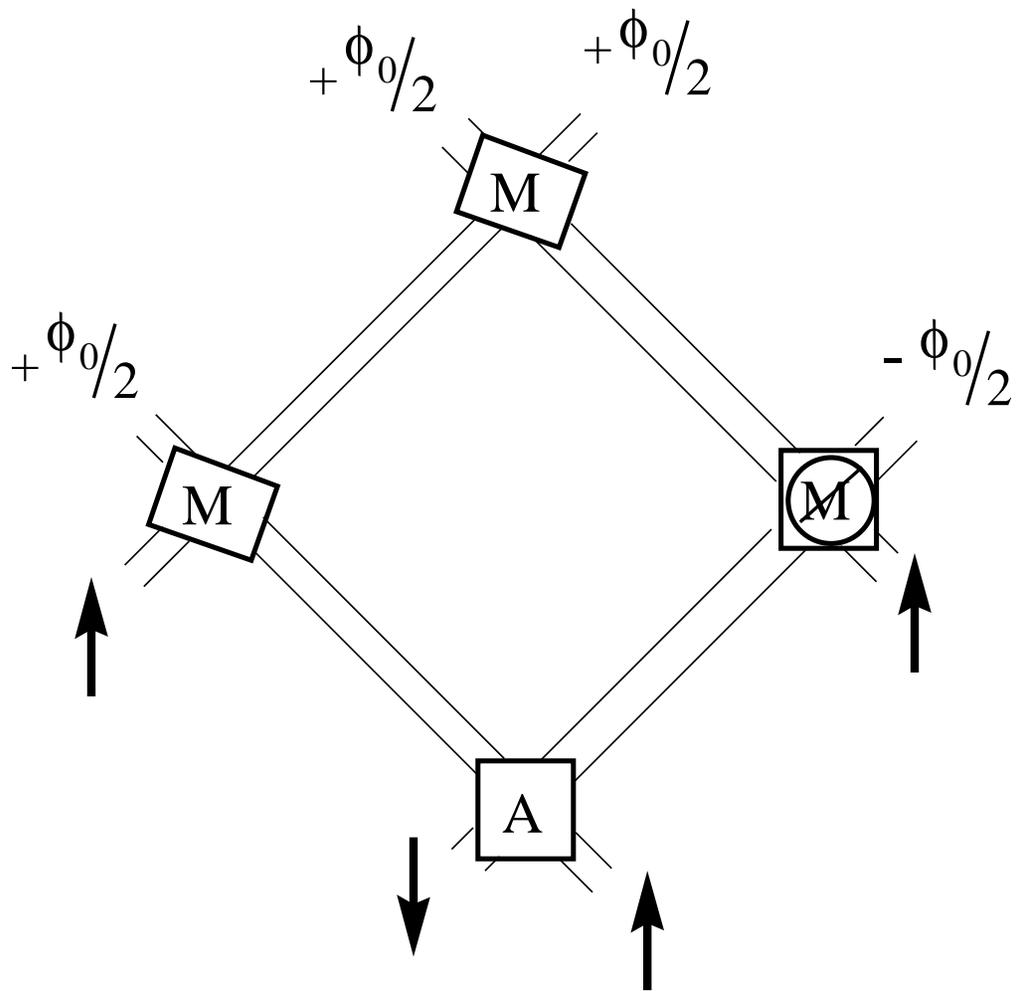
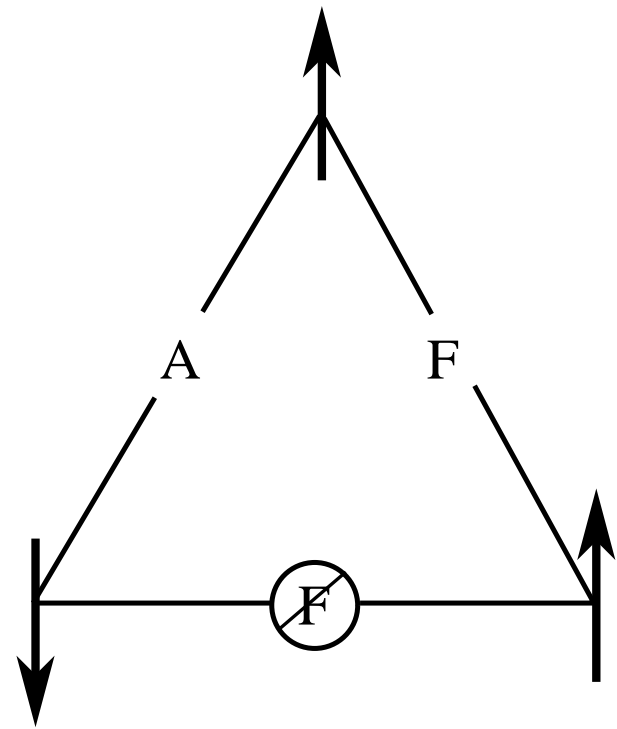
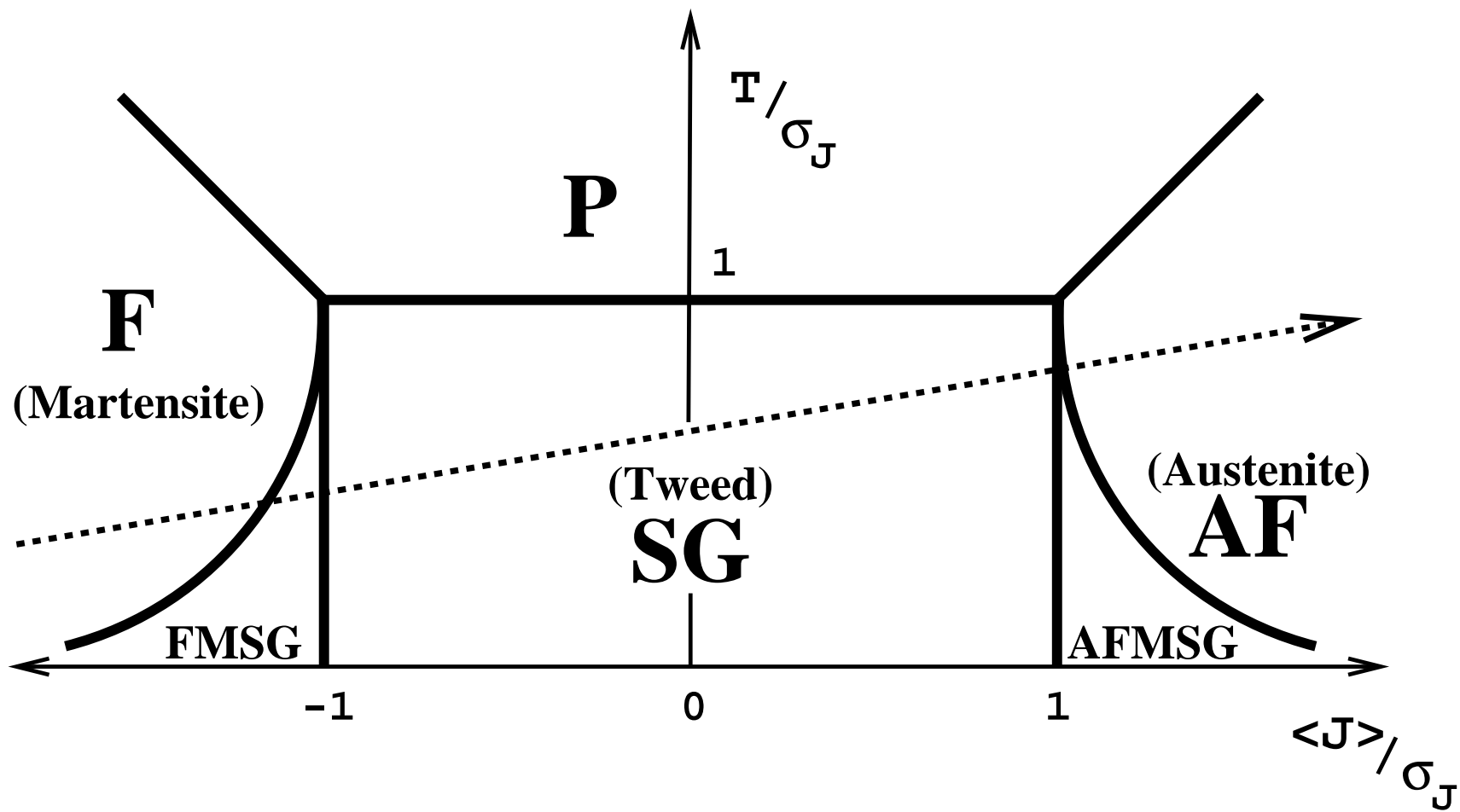
\end